\renewcommand{\[}{\begin{equation}}
\renewcommand{\]}{\end{equation}}
\def\spose#1{\hbox to 0pt{#1\hss}}
\def\lta{\mathrel{\spose{\lower 3pt\hbox{$\mathchar"218$}}
     \raise 2.0pt\hbox{$\mathchar"13C$}}}
\def\gta{\mathrel{\spose{\lower 3pt\hbox{$\mathchar"218$}}
     \raise 2.0pt\hbox{$\mathchar"13E$}}}

\let\boldgrk=\gkvecten
\let\boldgrksc=\gkvecseven

\def\gkthing#1{{\mathchoice%
	{\hbox{{\boldgrk\char#1}}}
	{\hbox{{\boldgrk\char#1}}}
	{\hbox{{\boldgrksc\char#1}}}
	{\hbox{{\boldgrksc\char#1}}}}}

\def\vtheta{\gkthing{18}}

\def\vmu{\gkthing{22}}

{\newif\ifnotend
\notendtrue
\def\veclist{ABCDEFGHIJKLMNOPQRSTUVWXYZabcdefghijklmnopqrstuvwxyz.}
\def\top#1#2.{#1}
\def\tail#1#2.{#2.}
\loop\expandafter\xdef\csname v\expandafter\top\veclist\endcsname%
{{\noexpand\bf\expandafter\top\veclist}}
\edef\veclist{\expandafter\tail\veclist}
\if\veclist.\notendfalse\fi\ifnotend\repeat}
 
\def\df{{\sc df}}\def\edf{{\sc edf}}

\def\pa{\partial}

\def\Rc{R_{\rm c}}

\def\kms{\,{\rm km}\,{\rm s}^{-1}}

\def\Gyr{\,{\rm Gyr}}
 
\def\pc{\,{\rm pc}}
\def\kpc{\,{\rm kpc}}

\def\e{{\rm e}}
\def\d{{\rm d}}

\def\feh{\hbox{[Fe/H]}}

\def\figref#1{Fig.~\ref{#1}}

\documentclass{iaus}
\usepackage{graphicx,natbib}

\title[Dynamical models and Galaxy surveys] 
{Dynamical models and Galaxy surveys}

\author[James Binney \& Jason Sanders]   
{James Binney$^1$ \& Jason L. Sanders$^1$
}

\affiliation{$^1$ Rudolf Peierls Centre for Theoretical Physics, Keble
Road, Oxford, OX1 3NP, UK\\ email: {\tt binney@thphys.ox.ac.uk},
{\tt jason.sanders@physics.ox.ac.uk}
}

\pubyear{2014}
\volume{298}  
\pagerange{XX -- YY}
\jname{IAUS\,298 Setting the scene for Gaia and LAMOST}
\editors{S. Feltzing, G. Zhao, N.\,A. Walton \& P.\,A. Whitelock, eds.}
\begin{document}

\maketitle

\begin{abstract}
Equilibrium dynamical models are essential tools for extracting science from
surveys of our Galaxy.
We show how models can be tested with data from a survey before the survey's
selection function has been determined. We
illustrate the application of this method by presenting some results for the
RAVE survey. We extend our published analytic distribution functions to
include chemistry and fit the chosen functional form to a
combination of the Geneva--Copenhagen survey (GCS) and a sample of G-dwarfs
observed at $z\sim1.75\kpc$ by the SEGUE survey. By including solid dynamics
we are able to predict the contribution that the thick disc/halo stars
surveyed by SEGUE should make to the GCS survey. We show that the measured
[Fe/H] distribution from the GCS includes many fewer stars at $\feh<-0.6$
than are predicted. The problem is more likely to lie in discordant abundance
scales than with incorrect dynamics.

\end{abstract}

\firstsection 
\section{Introduction}

Very significant resources are being invested in large surveys of the stellar
content of our Galaxy. We clearly are under an obligation to make every
effort to extract as much science from these expensive data as we can.  A
survey catalogue is always dominated by selection effects: it contains stars
that are nearby and luminous. Moreover several of the quantities measured for
stars, for example parallax $\varpi$, proper motion $\mu$, $\alpha$
enhancement $[\alpha/\hbox{Fe}]$, contain non-negligible errors, and the
quantities of physical interest that we derive from them, such as distance
$s$, luminosity $L$ and velocity $\vv$, have correlated errors with highly
non-Gaussian error distributions. Models play a vital role in extracting
science from surveys by taking into account (i) selection effects (ii)
measurement errors and (iii) combining constraints on the properties of the
Galaxy from different surveys. Here we illustrate these principles.

\section{Importance of equilibrium models}

There is a long tradition of extracting science from surveys with kinematic
models such as the Besan\c con model \citep{Robin03}. These models are constructed by assigning to each spatial point a
velocity ellipsoid, generally assumed to be Gaussian.  Equilibrium dynamical
models are superior to kinematic models in several ways.

\begin{itemize}

\item Although the Galaxy cannot be in dynamical equilibrium, equilibrium
models are of fundamental importance because an estimate of the Galaxy's
gravitational potential $\Phi(\vx)$ is key for any modelling
enterprise and we can constrain $\Phi$ only to the extent that the Galaxy is
in dynamical equilibrium: if it is allowed to be out of equilibrium, {\it
any} distribution of stars in phase space is consistent with {\it any}
gravitational potential -- we constrain $\Phi$ by assuming that the potential
is deep enough to prevent stars flying apart in the next dynamical time but
not so deep as to cause them to slump into the centre in the same period.

\item Equilibrium dynamical models have much less freedom than kinematic or
non-equilibrium models because their velocity structure is firmly tied to the
density structure by the assumed potential. Consequently a small number of
parameters suffices to specify a realistic dynamical model, and optimising
the model by adjusting these parameters is reasonably straightforward.

\item Observed velocity distributions are far from Gaussian but are well
reproduced by quasi-isothermal \df s.

\item A properly constructed equilibrium dynamical model can be used as the
foundation for the construction of non-equilibrium models via perturbation
theory. In this way the model can be extended to include non-equilibrium
phenomena such as spiral structure and the warp.

\end{itemize}

\section{Integrals of motion}

We have known since the numerical experiments of the 1960s that most
orbits in Galaxy-like flattened, axisymmetric potentials admit three
integrals of motion, and by the Strong Jeans Theorem the distribution
function (\df) of an equilibrium model can be assumed to be a function
$f(I_1,I_2,I_3)$ of these integrals. Since any function $J(\vI)$ of the
integrals is obviously itself an integral of motion, there is considerable
choice in what we use for arguments of the \df. One choice stands out above
all others: the action integrals $\vJ$. The special properties enjoyed by the
actions include: (i) they alone can be complemented by canonically conjugate
variables $\theta_i$ to form a complete set of canonical coordinates
$(\vtheta,\vJ)$ for phase space; (ii) the space in which the actions are
Cartesian coordinates, action space, provides an undistorted compression of
six-dimensional phase space in the sense that the volume of phase space
occupied by orbits with actions in $\d^3\vJ$ is $(2\pi)^3\d^3\vJ$; (iii)
actions are adiabatic invariants so when the Galaxy's potential is slowly
deepened, for example by the accretion of gas to form the disc, the orbits of
stars change but their actions do not. The property of adiabatic invariance
enables us to compute the effects of accretion without knowing precisely how
the accretion occurred -- we need to know only the initial and final
potentials, not the intermediate potentials. The ability to embed the actions
in a complete system of canonical coordinates is the key handling
non-equilibrium aspects of the Galaxy through perturbation theory --
angle-action coordinates were invented in the 19th century in order to compute
the dynamics of the solar system using perturbation theory. Particle physics,
condensed-matter physics, plasma physics, and celestial mechanics are all
built around perturbation theory, which is not merely the means by which we
compute the implications of models but provides the concepts -- Feynman
diagrams, phonons, spin-waves, mean-motion resonances, etc., -- with which we
understand the phenomena. Galactic dynamics will remain a primitive branch of
physics until it too makes extensive use of perturbation theory, and our
first step towards that goal must be expressing equilibrium models in terms
of angle-action coordinates.

The angular momentum around the Galaxy's approximate symmetry axis $L_z$ is
one of the three actions. The best expressions we have
\citep{Binney10,Binney12a,Sanders12} for the integral of motion that controls
the extent of a star's excursions perpendicular to the Galactic plane are for
the vertical action $J_z$. Consequently the only real temptation to employ an
integral that is not an action is the temptation to use the energy $E$ as the
third integral rather than the radial action $J_r$, which controls the extent
of a star's radial oscillation. For some the arguments advanced above for
choosing $J_r$ over $E$ do not suffice to wean them off $E$, and to these
people we say that if your \df\ has the form $f(E,L_z)$ it is quite
tricky to find the self-consistent gravitational potential
\citep{PrendergastT,Rowley}. Indeed, the search for $\Phi$ proceeds iteratively --
one guesses $\Phi$, evaluates the density $\rho=\int\d^3\vv\,f$ then implied,
solves Poisson's equation for the corresponding potential $\Phi'$ and repeats
the process until convergence. At each iteration the central potential
changes, with the result that the density depends upon a different range of
values of $E$, and the iterations converge only if some subtle scalings are
employed \citep[e.g.,][\S4.4.2(b)]{BT08}. Since actions always range from 0 for a
star that rests at the bottom of the potential well to $\infty$ for marginally
bound stars, when we adopt $f(\vJ)$ the iterations converge automatically.

In the following we take the \df\ to have the form $f(\vJ)$ and not waste
time on more traditional formulations.

\section{Choice of the DF}

\subsection{Basic DFs}

\cite{Binney10,Binney12b}  showed that superpositions of
quasi-isothermal \df s give good fits to the Geneva-Copenhagen Catalogue and
successfully predict data from the SDSS and RAVE surveys. A quasi-isothermal
\df\ has the form
 \[
f_{\rm iso}(\vJ)\propto\exp(-\Rc/R_\d)\exp(-\kappa J_r/\sigma_r^2)\exp(-\nu
J_z/\sigma_z^2),
\]
 where $R_\d$ is essentially the radial scale length of the disc, $\Rc(L_z)$
is the radius of a circular orbit of angular momentum $L_z$, $\kappa(L_z)$
and $\nu(L_z)$ are the radial and vertical epicycle frequencies of this
orbit, and $\sigma_r(L_z)$ and $\sigma_z(L_z)$ are approximately equal to the
radial and vertical velocity dispersions of the disc at $\Rc$.  The name
``quasi-isothermal'' for this \df\ derives from the observation that in the
epicycle approximation $\kappa J_r=E_r$ is the in-plane epicycle energy and
$\nu J_z=E_z$ is the energy of vertical oscillations, so the \df\ becomes an
exponential function of these energies like the Gibbs distribution of
standard statistical mechanics.

The Hipparcos data showed that the velocity dispersions of thin-disc stars
grow with age $\tau$ roughly as $\tau^{0.33}$ \citep[e.g.][]{AumerB}, so
\cite{Binney10} set
 \[
\sigma_i(L_z,\tau)=\sigma_{i}(L_z)
\left({\tau+\tau_1\over \tau_T+\tau_1}\right)^{0.33},
\]
 where $\tau_1$ is a parameter that controls the velocity dispersion of the
stars at birth, $\tau_T$ is the age of the oldest thin-disc stars and
$\sigma_{i}$ is essentially the present velocity dispersion at $\Rc$. This dispersion
is expected to decrease outwards through the disc, and an appropriate
functional form is
 \[
\sigma_{i}(L_z)=\sigma_{i0}\exp(-\Rc/R_\sigma),
\]
 where $\sigma_{i0}$ is a number, $R_\sigma$ is the radius over which the velocity dispersion falls by a
factor $\sim\e$. \cite{AumerB} showed that the Hipparcos data are consistent
with the hypothesis that the
star formation rate (SFR) near the Sun has varied with time as
$\exp(-t/\tau_f)$ so a reasonable form for the \df\ of the thin disc is
\[
f_{\rm thin}(\vJ)\propto\int\d\tau\,\exp(\tau/\tau_f)f_{\rm iso}(\vJ,\tau).
\]
 That is, the thin disc is modelled as a superposition of quasi-isothermal
discs, one for each coeval cohort of stars, with the velocity dispersion of
each cohort increasing with age. 

B12 took the thick disc to be a single quasi-isothermal
comprising exclusively old stars. The \df\ of the complete disc is then
 \[
f(\vJ)=(1-F)f_{\rm thin}(\vJ)+Ff_{\rm thick}(\vJ),
\] 
 where the individual \df s are normalised such that $\int\d^3\vJ\,f=1$ and
$F$ is the fraction of the mass of the disk contributed by the thick disc.

\subsection{Extended DFs}

In principle every observationally distinguishable class of star can have its
own \df.  Kinematics and chemistry are highly correlated, so stars with
different metallicities must have different \df s. We can model this
situation by extending the \df\ above to an extended distribution function (\edf),
which is a  function of chemical composition in addition to actions.

\begin{figure}[b]
\begin{center}
\centerline{\includegraphics[width=.4\hsize,angle=270]{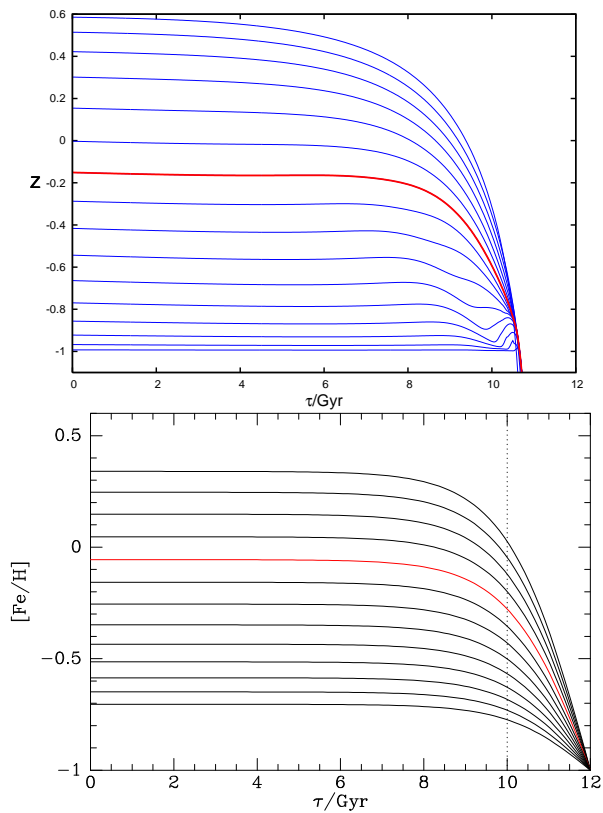}}
\vskip2pt
\centerline{\includegraphics[width=.58\hsize]{binney_f1b.ps}}
 \caption{The upper panel shows the evolution of $\feh$ in the ISM according
to the model of Sch\"onrich \& Binney (2009). Each curve corresponds to a
radius that increases from top to bottom in steps of $1.25\kpc$, the
curve for $R=8$ being shown in red. The lower panel shows the analytic
approximation to this evolution afforded by eqs.~(\ref{JJBeq:MRT}) and
(\ref{JJBeq:MR}). The increment in $R$ between curves is again $1.25\kpc$
with the curve for $R=8\kpc$ shown in red. The region to the right of the
dotted vertical line is taken to constitute the thick disc, which has the
\df\ of a single quasi-isothermal.}\label{JJBfig:Zhist}
\end{center}
\end{figure}

We start by supposing that the chemical composition of the interstellar
medium (ISM) is at any given time a function of Galactocentric radius $R$
only, and that the chemical composition of a star is the same as that of the
ISM at the coordinates $(R,\tau)$ of its birth. The upper panel of
Fig.~\ref{JJBfig:Zhist} shows the time dependence of metallicity at several
radii in the model of \cite{SchoenrichB}. We model this
dependence with
 \[\label{JJBeq:MRT}
\feh(R,\tau)=F(R,\tau)\equiv F(R)+[F(R)-F_m]
\left[\tanh\left(\tau_m-\tau\over\tau_f\right)-1\right],
\]
 where from SB09 we adopt the current metallicity-radius
relation 
\[\label{JJBeq:MR}
F(R)=\tanh\left\{0.6-0.082{R\over\!\kpc}\right\},
\]
 which yields a linear decline of [Fe/H] near the sun flattening to $-1$ at
large $R$. Here $\tau_m$ is the maximum age of any star in the Galaxy, $F_m$ is that
star's value of $\feh$, and $\tau_F$ is a parameter that controls the rate of
enrichment at early times. Following SB09 we adopt $\tau_m=12\Gyr$ and
$F_m=-1$. The lower panel of Fig.~\ref{JJBfig:Zhist} shows a reasonable fit
to the SB09 model is afforded by $\tau_F=1.98\Gyr$.

With the assumption that a star's metallicity is given by equation
(\ref{JJBeq:MRT}) with $R$ interpreted as $\Rc(L_z)$ the \edf\ becomes
 \[
f(\vJ,\feh)=\int\d\tau\,\exp(\tau/\tau_f) f_{\rm iso}(\vJ,\tau)\delta(\feh-F(\Rc,\tau)).
\]
 On account of churning
\citep{SellwoodB}, the present angular momentum $L_z$ of a star differs
from its birth angular momentum $L_z'$ by a random offset, which we assume is drawn from a Gaussian
distribution with dispersion $\sigma_L$ that grows with time, so
\[\label{JJBeq:defssigmaL}
\sigma_L(\tau)=\sigma_{L0}\left({\tau\over\tau_m}\right)^{\gamma_T}.
\]
 With churning taken into account,  the \edf\ becomes
\[\label{JJBeq:edf}
f(\vJ,\feh)=\int\d L_z'\,\int\d\tau\,\exp(\tau/\tau_f)
{\e^{-(L_z-L_z')^2/2\sigma_L^2}\over\sqrt{2\pi\sigma_L^2}}
 f_{\rm iso}(\vJ',\tau)\delta(\feh-F(\Rc',\tau)),
\]
 where $\vJ'\equiv(J_r,L_z',J_z)$ and $\Rc'\equiv \Rc(L_z')$. When we use the
 $\delta$-function to execute the integral over $L_z'$ we obtain
\[
f(\vJ,\feh)=\int\d\tau\,\exp(\tau/\tau_f)
{\e^{-(L_z-L_z')^2/2\sigma_L^2}\over\sqrt{2\pi\sigma_L^2}}
 {f_{\rm iso}(\vJ',\tau)\over |\pa F/\pa \Rc||\pa \Rc/\pa L_z|},
\] 
 where $L_z'$ is given by $F(\Rc(L_z'),\tau)=\feh$.

\section{Fitting an incomplete catalogue}

To test a model by comparing it with a catalogue, we have to
``observe'' the model in in the same way that the survey observes the Galaxy.
Most of the major surveys now underway make strenuous efforts to be
photometrically complete in some well-defined way. Nevertheless this
completeness is achieved, if at all, only in the final data release because a
host of complex constraints and considerations affect the order in which
targets are selected -- bright stars are likely to be selected ahead of faint
ones, and in crowded fields it may be impossible to sweep up adjacent targets
in the first visit to the field. Fortunately, these considerations are almost
always velocity-blind in the sense that the probability that a given star is
included in a particular data release is rigorously independent of its
velocity. In these circumstances it is easy to test a model with a catalogue
for which we have estimates of distance and chemical composition
as follows.

We bin the stars spatially and chemically in any convenient way.  
For  each star in a bin we choose a hypothetical true distance $s'$ using
either a simple Gaussian distribution in $s'$ or, more rigorously, using the
a posteriori probability distribution of $s'$ for each star that is returned
by a Bayesian distance-determination algorithm \citep{BurnettB,BinneyRAVE}.
The probability distributions from a Bayesian algorithm are clearly to be
preferred because they take into account the higher density of stars
near us than further away and also distance ambiguities associated with
uncertainty as to whether a star is a dwarf or some type of giant.
In the same spirit we select a hypothetical true metallicity $\feh'$
by sampling the product of the model's metallicity distribution along that line
of sight and the probability distribution of measurement errors in $\feh$.
Once hypothetical true values of all relevant observables have been selected
in this way, we sample the model's velocity distribution at the hypothesised
location
 \[
P(\vv)=f(\vJ(\vx',\vv),\feh',\ldots).
\]
 We convert $\vv$ into the line of sight velocity $v_\parallel$ and proper
motion $\vmu$ using the hypothesised distance $s'$. Next we add to
$v_\parallel$ and $\vmu$ appropriate measurement errors before converting
them back to a space velocity $\vv$ using the measured distance $s$.  These
velocities constitute the model's predictions for the distribution of
velocities of catalogued stars in the given spatial bin. These predictions
take fully into account all measurement errors, no matter how gross, in
distances, velocities and metal abundances. If the measurement errors have
been correctly assessed, any statistically significant discrepancy between
this theoretical velocity distribution and the observed one must reflect a
shortcoming of the model.

\section{Metallicity-blind predictions from the GCS catalogue}

B12 fitted a \df\ $f(\vJ)$ to the Geneva-Copenhagen Catalogue (GCS) of F and
G stars \citep{Nordstrom,Holmberg} as follows. First the thin-disc \df\ was
fitted to the histograms of $U$, $V$ and $W$ velocity components. Then the
data set was augmented by the \cite{GilmoreR} estimates of
the stellar density $\rho(z)$ at distance $z$ from the plane, and the entire
\df\ was fitted.  Since the fits to all the data, especially the distribution
of $W$ components and the GR83 density points, were excellent, it was evident
that the gravitational potential adopted is not far from the truth. However,
more rigorous tests of the (\df,$\Phi$) combination of this model are
provided by using it to predict the velocity distributions of RAVE stars,
which, unlike the GCS stars, extend far beyond a sphere around the Sun of
radius $\sim150\pc$.

\begin{figure}
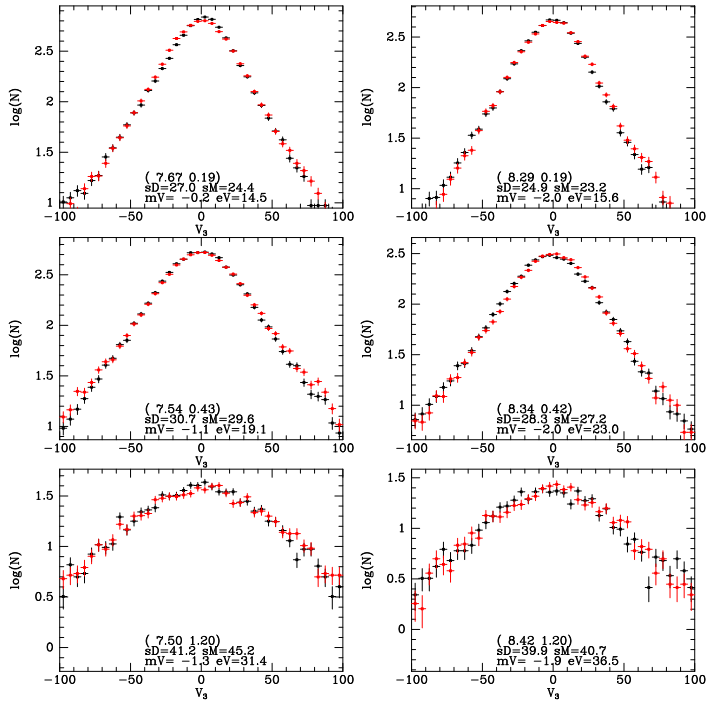

\centerline{
\includegraphics[width=1.8in]{binney_f2a.ps}
\includegraphics[width=1.8in]{binney_f2b.ps}}
\centerline{
\includegraphics[width=1.8in]{binney_f2c.ps}
\includegraphics[width=1.8in]{binney_f2d.ps}}
\centerline{
\includegraphics[width=1.8in]{binney_f2e.ps}
\includegraphics[width=1.8in]{binney_f2f.ps}}
 \caption{Vertical velocity components of giant stars in RAVE. Left column:
stars with $R<R_0$, right column $R>R_0$. The red points are the prediction
of the model for thin-disc stars, the black points show velocities measured
by RAVE. The numbers in brackets at the bottom give the mean values of $R$
and $|z|$ for stars in that bin.}\label{JJBfig:gW}
\end{figure}

\begin{figure}
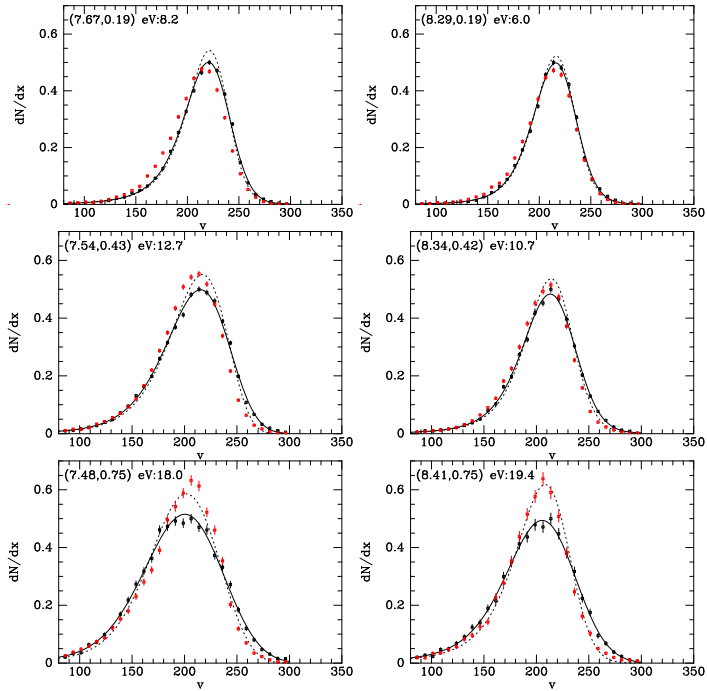

\centerline{
\includegraphics[width=1.8in]{binney_f3a.ps}
\includegraphics[width=1.8in]{binney_f3b.ps}}
\centerline{
\includegraphics[width=1.8in]{binney_f3c.ps}
\includegraphics[width=1.8in]{binney_f3d.ps}}
\centerline{
\includegraphics[width=1.8in]{binney_f3e.ps}
\includegraphics[width=1.8in]{binney_f3f.ps}}
\caption{$v_\phi$ components of RAVE giants. Left column: stars with $R<R_0$,
right column $R>R_0$. The red points are the prediction of the model, the
black points show velocities measured by RAVE. The numbers in brackets at top
left give the mean values of $R$ and $|z|$ for stars in that bin followed by
the mean observational error. The dotted curves show models of the underlying
error-corrected distribution and the black curves show the result of
convolving these distributions with the stated errors.}\label{JJBfig:gvphi}
\end{figure}

Figs.~\ref{JJBfig:gW} and \ref{JJBfig:gvphi} show results for giants. We
divide the stars into those inside/outside the solar radius, and into three
bins in $|z|$. The black points show histograms for $\sim140\,000$ RAVE
giants; each point's  vertical bar shows the statistical error in that bin.
The red points show the corresponding theoretical predictions.

\figref{JJBfig:gW} shows histograms for the vertical velocity component $V_3$ (the
direction of this component changes from point to point in the $(R,z)$ plane
to track a principal axis of the velocity ellipsoid). The mean
$(R,|z|)$ coordinates of the contributing stars appear in brackets in the
lower centre of each panel above the error-corrected velocity dispersion in
the bin (sD) and the mean velocity (mV).  In all panels the red and black
histograms agree nearly perfectly, even though the stars that contribute to the
lower four panels lie far outside the region occupied by the GCS stars, to
which the \df\ was fitted.

\figref{JJBfig:gvphi} shows histograms for $v_\phi$. Although not perfect,
the agreement between the red and black histograms is again impressive given
that the red points are predictions rather than fits to these data. The fit
becomes near-perfect if one assumes that the distances to the most remote
stars have been overestimated by $\sim20\%$.

\section{Including metallicity information}

In the last section we merely tested against RAVE data a previously selected
basic \df. Now we report fitting an \edf\ to a combination of SEGUE and GCS
data. In this work we include a halo described by a classical isothermal \df\
$f\propto\exp(-E/\sigma^2)$. The parameters of the halo, including the local
density of its stars, are fixed. Its metallicity distribution is a Gaussian
centred on $\feh=-1.5$ with dispersion $0.5$. It contributes $\sim0.1\%$ of
local stars.

The thick disc was taken to be a single quasi-isothermal distribution with
the metallicity-angular momentum relation that one gets by integrating with
respect to $\tau$ within the lower panel of \figref{JJBfig:Zhist} from
$\tau=10\Gyr$ to $12\Gyr$.  Likelihood maximisation was first used to fit the
\edf\ of the thick disc to G dwarfs in the DR9 SEGUE catalogue with $R>R_0$
and $z$ in $(1.5,2)\kpc$.   $N\simeq50$ possible values of $\vx$, $\vv$ and
$\feh$ were chosen for each star $\alpha$ from that star's error
distribution. For each such coordinate set the ratio
 \[\label{JJBeq:defsn}
n(\vv_j)\equiv {f_{\rm thick}(\vx_j,\vv_j,\feh_j)\over
\int\d^3\vv\,f_{\rm thick}(\vx_j,\vv,\feh_j)}
\]
 is the normalised probability density of the assigned velocity given the
position and metallicity assignments. Crucially, by dividing by the velocity
integral we have made $n(\vv_j)$ insensitive to the spatial and metallicity
coordinates, and therefore to the survey's selection function.  The average
 \[
p_\alpha\equiv {1\over N}\sum_{j=1}^Nn(\vv_j)
\]
 over all randomly sampled velocities $\vv_j$ is a statistically more stable
measure of the probability of the star's measured velocity (around which the
$\vv_j$ cluster) that inherits the same insensitivity to the selection
function. In order to ensure that the density profile $\rho(z)$ of the thick
disc is appropriate, the parameters of the thick-disc \edf\ were chosen by
maximising the function
 \[\label{JJBeq:logL}
{\cal L}=\sum_{\rm stars\ \alpha}p_\alpha-
\sum_{z>1.3\kpc}\left|{\log_{10}[\rho_{\rm GR}(z)/\rho_{\rm
DF}(z)]\over\sigma_{\rm GR}(z)}\right|^2,
\]
 where $\rho_{\rm GR}$ is the density profile from GR83, $\sigma_{\rm GR}$
are the errors in $\log_{10}(\rho_{\rm GR})$, and $_\rho{\rm DF}$ is the
density profile predicted by the \edf.  

After pinning down the parameters of the thick-disc \edf\ in this way, the
GCS velocities, with $\feh$ values from \cite{Casagrande11} were used to
maximise a log likelihood similar to equation (\ref{JJBeq:logL}) but (a)
using all the GR83 points in the sum over $z$, and (b) with $n$ in
eq.~(\ref{JJBeq:defsn}) made
sensitive to $\feh$ in addition to $\vv$ by adding integration over $\feh$ to
the denominator
 \[
n(\vv_j)\equiv {f_{\rm all}(\vx_j,\vv_j,\feh_j)\over
\int\d^3\vv\int\feh\,f_{\rm all}(\vx_j,\vv,\feh)},
\]
 where $f_{\rm all}$ is the complete \edf\ rather than just of just that of
the thick disc. 

\begin{figure}
\centerline{\includegraphics[width=.8\hsize]{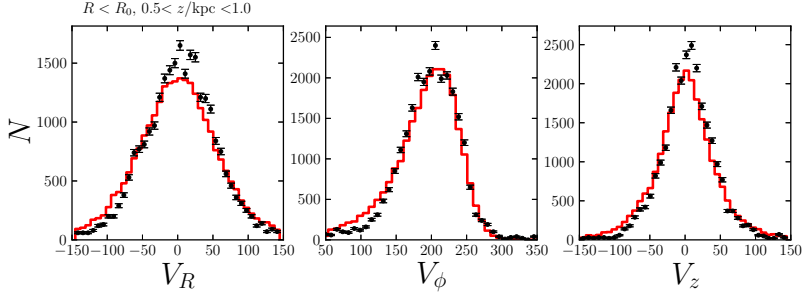}}
\caption{Red histograms: predicted velocity distributions for SEGUE G dwarfs
in the spatial bin $R<R_0,\, 0.5<|z|<1$.
}\label{JJBfig:SEGUEdw}
\end{figure}

\begin{figure}
\centerline{\includegraphics[width=.85\hsize]{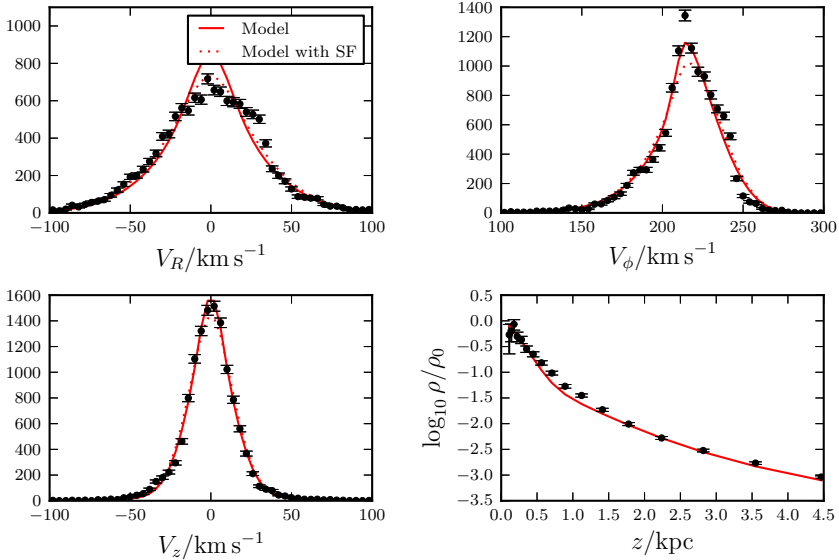}}
\caption{Fits of the model to the three velocity histograms of GCS stars and
the Gilmore-Reid density profile.}\label{JJBfig:GCSrho}
\end{figure}
\begin{figure}
\centerline{\includegraphics[width=.55\hsize]{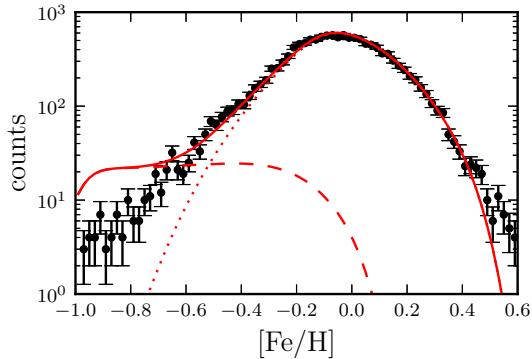}}
\caption{Data points: the metallicity distribution of the GCS from
\cite{Casagrande11}. Red curve the distribution predicted by fitting the
thick disc to the SEGUE G-dwarf distribution and then adding the thin disc.
The red dashed curve shows the matallicity distribution of the SEGUE
stars.}\label{JJBfig:SEGUEFe}
\end{figure}

\figref{JJBfig:SEGUEdw} shows that the final model provides reasonable if not
spectacular fits to the velocities of SEGUE dwarfs that were not used in the
fitting procedure. The lower-left panel of \figref{JJBfig:GCSrho} shows that
the model provides an excellent fit to the distribution of $W$ components of
GCS stars, while the upper two panels show that the fits to the $U$ and $V$
components are imperfect at low velocities, presumably on account of the
prominence of streams near the origin of the $(U,V)$ plane \citep{Dehnen98}.

Our fitting procedure yields the joint probability distributions of  the
parameters of each disc. We do not have space to describe these distributions
fully here, but we note a few points:

\begin{itemize}  

\item The thick disc is hotter vertically than horizontally:
$43\kms\simeq\sigma_{r0}<\sigma_{z0}\simeq53\kms$. The velocity dispersion
parameters of the thin disc are as expected
($37\kms\simeq\sigma_{r0}>\sigma_{z0}\simeq23\kms$).

\item The thick-disc \edf\ has a small scale length parameter
($R_\d\lta2.5\kpc$).  However, this parameter describes the distribution of
birth radii of stars, and on account of strong radial migration described by
the parameter $\sigma_L$ in equation (\ref{JJBeq:defssigmaL}), the present
distribution of thick-disc stars would be described by a significantly larger
scale length $\sim3.5\kpc$.

\item The scale length parameter of the thin disc is larger than that of the
thick disc ($\sim3.5\kpc$). The spatial distribution of thin-disc stars is
broadened by radial migration, even if not to the same extent as in the thick
disc, so the current scale length is at the upper end of former estimates.
The thin disc has a remarkably large value, $R_\sigma\gta15\kpc$, of the scale
length on which the dispersions decrease radially. 

\item The thick disc contributes $\sim9\%$ of local stars, a value that is
smaller than those ($\sim13\%$ and $20\%$) estimated by \cite{Juric08} and
\cite{Fuhrmann11} but a factor of 3
larger the the original estimate of GR83. Given that the model density profile
$\rho(z)$  falls below the GR83 points in the range $0.5<|z|/\!\kpc<1.5$
(lower right-hand panel of \figref{JJBfig:GCSrho}), the normalisation of the
thick disc may be too low. However,  \figref{JJBfig:SEGUEFe}
shows that the metallicity distribution of the GCS stars provide a contrary
indication:  when the thick-disc normalisation is large
enough to fit the GR83 points, the GCS metallicity distribution shows fewer
metal-poor stars than the SEGUE data require.  \cite{Juric08} derived a
profile $\rho(z)$ from the SDSS data that is slightly less steep than the
GR83 data, so using this profile would raise the normalisation of the thick
disc's contribution to the \edf\ and thus exacerbate the conflict with the
GCS metallicity distribution.

\end{itemize}

How serious is is our inability to fit the metal-weak tail of the GCS
metallicity distribution simultaneously with the GR83 density profile and the
SEGUE metallicity distribution? Possible explanations include: (i) One or
both of the GCS and SEGUE metallicity scales is wrong; (ii) During the
formation of the thick disc, the star-formation rate varied rapidly, with the
consequence that the actual metallicity distribution in the thick disc is not
adequately approximated by the one we have gleaned from the right-hand end of
the lower panel of \figref{JJBfig:Zhist} under the assumption that of a
constant SFR. (iii) The thick disc's \df\ is not approximately
quasi-isothermal. Specifically, if the \df\ were not a monotonically
decreasing function of $J_z$, it would be possible to have the required
number of metal-poor stars at $|z|\sim1.75\kpc$ without exceeding the number
seen at $z\sim0$ because the density profile $\rho(z)$ of the thick disc
could be non-decreasing at $|z|\lta1\kpc$. Two strong objections can be
raised to this fix: (a) the proposed density profile conflicts with the
conclusion of \cite{Bovy12} that mono-abundance populations have
double-exponential density profiles, as predicted by quasi-isothermal \df s;
(b) if, as seems likely, the thick disc formed by stars being scattered from
near-circular orbits during an early disorderly phase in the life of the
Galaxy, the \df\ must decrease in the direction that stars diffused, namely
that of increasing $J_z$. For these reasons we consider it unlikely that the
conflict arises from the assumed form of the \edf\ and is more likely to
arise from either differences in metallicity scales or rapid early
fluctuation in the SFR.

\section{Conclusions}

Equilibrium dynamical models are crucial for the scientific exploitation of surveys. We
have given two examples of their use. First we have shown that a  \df\ fitted
to the velocity distribution of local stars plus the GR83 density profile
predicts the velocity distributions of RAVE stars with
remarkable success given that most RAVE stars lie far from the Sun. Second we
have introduced an analytic \edf\ to model the correlations between
kinematics and chemistry. The \edf\ is inspired by a particular picture of
disc formation but its validity is ultimately independent of its original
physical motivation. We fitted this \edf\ to a combination of the SEGUE data
for G dwarfs seen $\sim1.75\kpc$ from the plane and the local GCS stars. The
resulting \edf\ makes reasonably successful predictions for kinematic data
not used in the fitting process, but provides an unsatisfactory fit to the
distribution in [Fe/H] of the GCS stars in the sense of requiring more
low-metallicity stars than observed. We discussed possible resolutions of
this discrepancy, and none is very attractive.

The fitting process suggests that for the thick disc the ratio of
velocity dispersions
$\sigma_r/\sigma_z$ is less than unity, whereas for the thin disc it much
exceeds unity.

\begin{discussion}

\discuss{Reddy}{Your model shows clear separation of the thin and thick discs
at $10\Gyr$. I think the thick disc evolves beyond the ``knee'' and hence
there must be many metal-rich young thick-disc stars younger than $10\Gyr$}

\discuss{Binney}{Although our model involves age as a parameter that controls
dispersion and chemistry, we don't have -- and have no prospect of getting --
the observational ability to distinguish stars with ages say one Gyr either
side of our brutal division of the disc into thin and thick at $\tau_T$. So
in practice it would make no difference if there were overlap between the
ages of stars in the two discs.}

\discuss{Rix}{Why do you consider the present-day velocity dispersions of
disc (sub-) populations mostly (or exclusively) as a result of heating. Could
they in good part not be a ``birth property''?}

\discuss{Binney}{We know that much or all of the thin disc has acquired
random velocity through stochastic acceleration by fluctuations in the
gravitational field. So I submit that the correct procedure scientifically
is to assume that {\it all} disc stars were born on near circular orbits.
This assumption may be false, but I want the data to demonstrate that, which
they can do only if I make the assumption and show it leads to conflict with
the data.}

\discuss{Minchev}{1. You showed that the ratio of vertical to radial velocity
dispersions is larger for the thick disc stars. Could this be seen as an indication
that the Milky Way disc was affected by mergers at high redshift? 2. Is it
possible to predict the past evolutionary history of the MW disc using the DF
approach?}

\discuss{Binney}{1. Before we speculation as to how
the Galaxy got to its present configuration, we should establish beyond
reasonable doubt how it is {\it presently} configured and how it works. I
find it distracting to be drawn into speculation about origins at this
early stage. 2. The DF of an equilibrium model, by definition, has the same
dynamical past as future. Information about the past, if it lingers, is encoded in the
distribution of stars in angle space, which I'm deliberately assuming to be
uniform. In reality it will be non-uniform. The key point is that you cannot
hope to make progress with probing the past until you have achieved clarity
about the present.}

\discuss{Chen Yun-teng}{How to interpret discrepancy between observables and
models? How to choose to quantify the comparison of models and observations.}

\discuss{Binney}{Since we are currently comparing theoretical and
measured 1-d histograms, we could use either $\chi^2$ or the
Kolmogorov-Smirnov measures of discrepancy. Actually Jason Sanders is using
$\chi^2$ and I have used ``$\chi^2$ by eye''. Really we should be comparing
densities in full 3d velocity space, and for that I suppose only $\chi^2$ is
feasible.}

\discuss{Ritter}{The comparison to external data sets has shown that there is
pretty much no connection between RAVE $\alpha$ enhancements and $\alpha$
enhancements derived from high-resolution, high SNR  spectra. This could
explain some of the discrepancies between the RAVE data and the model. I
suggest you use the $\alpha$ enhancements from Boeche's element abundance
catalogue instead, which should be more reliable than the official RAVE
values.}

\discuss{Binney}{We {\it are} using the Boeche et al.\ values. 
All the results were
resticted to $\feh$.}
\end{discussion}

\end{document}